\documentclass[prl,twoside,twocolumn,showpacs]{revtex4-1}

\usepackage{epsfig,bm,epsf,graphics}
\usepackage{graphicx}
\usepackage{dcolumn}
\usepackage{bm}
\usepackage{color}
\usepackage[T1]{fontenc}

\newcommand{\dd}{\hbox{\rm d}}
\begin{document}

\title{The Universality of Cancer}

\author{Carson C Chow, Yanjun Li and Vipul Periwal}%
\email{Corresponding author: vipulp@niddk.nih.gov}
\address{Laboratory of Biological Modeling, NIDDK, NIH, 9000 Rockville Pike, Bethesda MD 20892}

\date{\today}

\begin{abstract}
Cancer has been characterized as a constellation of hundreds of diseases differing in underlying mutations and depending on cellular environments. Carcinogenesis as a stochastic physical process has been studied for over sixty years, but there is no accepted standard model. We show that the hazard rates of all cancers are  characterized by a simple dynamic stochastic process on a half-line, with a universal linear restoring force balancing a universal simple Brownian motion starting from a universal initial distribution. Only a critical radius defining the transition from normal to tumorigenic genomes distinguishes between different cancer types when time is measured in cell-cycle units. Reparametrizing to chronological time units introduces two additional parameters: the onset of cellular senescence with age and the time interval over which this cessation in replication takes place. This universality implies that there may exist a finite separation between normal cells and tumorigenic cells in all tissue types that may be a viable target for both early detection and preventive therapy.
\end{abstract}

\pacs{87.19.xj, 89.75.Fb, 87.14.gk}
 
 \maketitle


Cancer is considered to be a multifaceted disease where the phenotypic similarities of tumor progression are a veneer over a multitude of possible underlying genetic alterations~\cite{varmus}.  Three-quarters of all cancers are probably sporadic. As an organism ages, the accumulation of mutations increases the likelihood of alteration in an oncogene or in a tumor suppressor gene, which in turn can lead to an accumulation of mutations. The process of carcinogenesis has been modeled for over 60 years~\cite{nordling,armitage,cook41,mool,pw,ritter40,morganthaler42,arbeev26,gsteiger43,hplw,mdz38,gerstung39,tofigh}, but there is no consensus model. Summaries of the state of cancer incidence modeling can be found in Harding, Pompei and Wilson~\cite{hpw} and Beerenwinkel et al.~\cite{beeren}.

Recently, Tomasetti and Vogelstein~\cite{tv}
showed that the lifetime risk of cancers of many different types is correlated with the total number of divisions of the normal self-renewing cells, the somatic stem cells, that maintain each tissue's homeostasis. This implies that most cancer is due to random mutations arising during DNA replication in normal, noncancerous somatic stem cells. This observation poses an interesting challenge for a mechanistic understanding of cancer incidence. If a somatic stem cell can become cancerous at any time during a lifetime, any valid model of carcinogenesis should be able to use this as a basis for age-specific cancer hazard rate prediction and match available age-specific cancer incidence rates\cite{seer}. 

Harding et al.~\cite{hpw} have analyzed the Surveillance, Epidemiology and End Results (SEER\cite{seer}, specifically SEER 9) cancer registries to compile age-specific incidence rates, with particular care accorded to the data on the very elderly (ages > 80 years).  They noted that incidence for most cancers reached a maximum between ages 75 years and 90 years, with a precipitous decline later, and often tended to vanish among centenarians. This decline is difficult to explain in stem cell models that assume that any stem cell will eventually produce a tumor~\cite{gerstung39}.  With exceptions, for example, the beta model~\cite{pw} and the generalized beta model~\cite{mdz38}, such models project increasing cancer rates throughout adulthood. 

Our aim here is to present a simple universal physical model for the stochastic process of tumorigenesis that resolves the tension between the spontaneous random origin of cancer shown in Ref.~\cite{tv} and observed age-specific incidence rate curves~\cite{seer,hpw}. While tissues are heterogeneous in cellular characteristics, recent work on inducing pluripotency\cite{pluri} in differentiated cells\cite{escell}, along with work on somatic stem cells with regard to cancer\cite{gilbertson}, suggests that such stem cells share commonalities. If so, the fundamental process of cellular replication, and the fidelity of the concomitant information propagation, is most likely to be universal. From a physical perspective, the propagation of information by replication is a stochastic process with error correction in the form of a multitude of repair mechanisms\cite{dnadamage}. We expect, then, that a limit on error correction should be universal to all cancers in a given species. In other words, irrespective of tissue or cell type, we expect that there should be a coordinate, measuring the effective error in the propagated genome, that at a critical value marks a sharp transition between normal and tumor cells. If there is no such sharp transition, the utility and feasibility of early detection of cancer is called into question. Here, we show that that there is such a coordinate, and a universal diffusion process for the position of each stem cell on this coordinate, such that  age-specific cancer hazard rates are determined by the probability of crossing a cancer--type--independent error limit. Diffusion processes have been studied in this connection\cite{tom}, but the focus was on the accumulation of mutations in the pre-cancer phase. The role of error correction in computing cancer incidence rates has not been investigated.

We set out to compare possible diffusive processes for different tissues. Although DNA mutations and epigenetic changes can be introduced into a non--mitotic cell, (e.g. through a viral insertion or the action of retrotransposons), we focus on the errors that accumulate during cell division (i.e. S phase). 
The human genome has $3\times 10^9$ base pairs that can mutate but not all genome positions contribute equally to cancer susceptibility. We introduce an effective error coordinate, $r,$ parameterizing an appropriately weighted mean alteration distance in genome space, including both mutations and epigenetic changes, such as methylation changes that affect DNA repair genes themselves. We hypothesized that cancer occurs when this error coordinate exceeds a critical value, $r_c.$ 

The cell constantly expends energy for DNA error correction, which acts as a restoring force to oppose mutational diffusion~\cite{dnadamage}. We choose  a restoring force that scales linearly with error coordinate. The error coordinate therefore obeys an Ornstein-Uhlenbeck stochastic process~\cite{ou} on the half line. There are two physical motivations for this choice of restoring force: First, the error correction response is then tolerant of infinitesimal excursions away from the original genome, and second, a discrete urn model of mutating bases has this process as a scaling limit. 

The probability density for $r$ obeys the Fokker-Planck equation
\begin{equation}\label{eq:ou}
\partial_\tau p(r,\tau) = r_\theta^{-2} \partial_r rp(r,\tau) + \partial_r^2 p(r,\tau)
\label{fp}
\end{equation}
with a reflecting boundary condition at the origin (i.e. $\partial_rp(r=0,\tau)=0$).  This equation can be analytically solved. We allow for the possibility of mitotic mutations during prenatal development by parameterizing the initial density as $n(r,\tau=0) \propto \exp(-\alpha r^2/2r_\theta^2)$, where $\alpha$ is dimensionless and $\alpha = \infty$ for an error free initial genome.  Given that we only consider error accumulation during mitosis, the relevant time scale is measured in terms of the number of cell divisions, which may not be constant in time.  In fact, it is well known that cells become senescent with age.  Hence, to relate $\tau$ to chronological time $t$, we introduce a reparameterization 
\begin{equation}\label{eq:tau}
\dd\tau = \dd t\ D(t) \equiv \dd t\ 0.5 \bigg(1+\tanh\Big({T_s-t\over w_s}\Big)\bigg),
\end{equation}
where $T_s$ is a cell--type specific senescence time, and $w_s$ a cell--type specific senescence time uncertainty, motivated by Ref.~\cite{hpw}. The probability of having become cancerous is given by $p(r>r_c)=\int_{r_c}^\infty p(s,t) ds$. The cancer hazard rate is then the
 time derivative of the odds $p(r>r_c)/p(r<r_c)$:
 \begin{equation}\label{eq:ir}
I(t) = {\dd\over {\dd t}}{\int_{r_c}^\infty \dd s\ p(s,t)\over {\int_0^{r_c} \dd s\ p(s,t)}}
\end{equation}
Hazard rates are better estimated from incidence\cite{seer} rather than mortality data\cite{hpw}.

%
%
%


The model is specified by the parameters $\alpha$, $r_\theta$, $T_s$, $w_s$ and $r_c.$ Our goal is to fit the predicted cancer hazard rate to the observed incidence rates for various cancer types\cite{seer}.  However, since we are free to choose a scale to measure errors, we can fix $r_\theta=1.$ Using Nelder-Mead minimization for parameter determination, we only needed to fit the incidence curve $I(t)$ up to a scale factor since our theory considers the per--stem--cell hazard rate while the total hazard rate is this rate summed independently over all stem cells in the tissue~\cite{tv}. If there are only $x$ susceptible cells in the tissue, then the hazard rate for that tissue is  $x$ times the rate for a single somatic stem cell in that tissue.
\begin{table}
\caption{\label{tab:table1} {Parameters for various cancers (NS, Nervous System; (non-)Hodgkin, (non-)Hodgkin Lymphoma)}
}
\begin{ruledtabular}
\begin{tabular}{|cc|c|ccccc|}
&Tissue &Sex &BIC 
&$T_s (yrs)$ & $w_s (yrs) $ & $r_\alpha$ &$r_c$\\
\hline
&  All &  M &  418 &  93.7 &  14.2 &  0.119 &  2.68 \\ 
\hline
&  Brain \& NS  &  M &  45.6 &  90.3 &  7.15 &  0.259 &  2.5 \\ 
\hline
&  Breast &  M &  67.7 &  97.7 &  0.515 &  0.151 &  2.5 \\ 
\hline
&  Colon &  M &  178 &  98 &  7.32 &  0.247 &  2.88 \\ 
\hline
&  Esophagus &  M &  29.4 &  92.8 &  14.2 &  0.1 &  2.59 \\ 
\hline
&  Hodgkin1 &  M &  14.4 &  6.18e-08 &  32.8 &  0.159 &  1.19 \\ 
\hline
&  Hodgkin2 &  M &  129 &  90.2 &  12.7 &  0.1 &  2.25 \\ 
\hline
&  Kidney &  M &  1e+04 &  92.3 &  13 &  0.1 &  2.38 \\ 
\hline
&  Larynx &  M &  110 &  80.6 &  16.3 &  0.166 &  2.86 \\ 
\hline
&  Leukemia &  M &  342 &  103 &  2.7 &  0.269 &  2.83 \\ 
\hline
&  Liver &  M &  290 &  94.6 &  9.4 &  0.199 &  2.34 \\ 
\hline
&  Lung &  M &  120 &  88.5 &  14.8 &  0.181 &  3.02 \\ 
\hline
&  Melanoma &  M &  150 &  97.5 &  5.01 &  0.229 &  2.29 \\ 
\hline
&  Mesothelioma &  M &  50.2 &  87.4 &  7.33 &  0.285 &  3.97 \\ 
\hline
&  Misc &  M &  80.3 &  99.5 &  3.93 &  0.219 &  3.08 \\ 
\hline
&  Myeloma &  M &  18.5 &  96 &  9.22 &  0.183 &  2.65 \\ 
\hline
&  Non-Hodgkin &  M &  76.2 &  97.1 &  9.67 &  0.194 &  2.57 \\ 
\hline
&  Oral &  M &  501 &  98.5 &  7.86 &  0.1 &  1.98 \\ 
\hline
&  Pancreas &  M &  22.8 &  96.9 &  7.52 &  0.261 &  2.89 \\ 
\hline
&  Prostate &  M &  2.48e+03 &  74.7 &  22.9 &  0.1 &  3.51 \\ 
\hline
&  Stomach &  M &  33.5 &  99.5 &  5.95 &  0.256 &  2.88 \\ 
\hline
&  Testis &  M &  33.4 &  28.9 &  20.1 &  0.135 &  1.59 \\ 
\hline
&  Thyroid &  M &  23.9 &  87.2 &  12.2 &  0.242 &  1.9 \\ 
\hline
&  Urinary &  M &  2.68e+03 &  97.2 &  9.04 &  0.182 &  2.89 \\ 
\hline
&  All &  F &  2.01e+03 &  99.9 &  13.4 &  0.197 &  2.4 \\ 
\hline
&  Brain \& NS &  F &  38.1 &  92.4 &  6.85 &  0.327 &  2.5 \\ 
\hline
&  Breast &  F &  1.64e+03 &  91 &  30.3 &  0.1 &  2.2 \\ 
\hline
&  Cervix Uteri &  F &  204 &  96.7 &  7.66 &  0.103 &  1.33 \\ 
\hline
&  Colon &  F &  51.6 &  100 &  10.4 &  0.228 &  2.91 \\ 
\hline
&  Corpus Uteri &  F &  516 &  81.1 &  19.1 &  0.1 &  2.43 \\ 
\hline
&  Esophagus &  F &  18.4 &  99.3 &  5.79 &  0.282 &  2.87 \\ 
\hline
&  Hodgkin1 &  F &  13.4 &  3.12e-09 &  22.4 &  0.142 &  1.37 \\ 
\hline
&  Hodgkin2 &  F &  124 &  93 &  12.1 &  0.1 &  2.19 \\ 
\hline
&  Kidney &  F &  50.9 &  93 &  11 &  0.214 &  2.46 \\ 
\hline
&  Leukemia &  F &  192 &  102 &  4.63 &  0.308 &  2.9 \\ 
\hline
&  Lung &  F &  36.8 &  82.7 &  14 &  0.12 &  3.08 \\ 
\hline
&  Misc &  F &  57.2 &  107 &  5.01 &  0.428 &  3.5 \\ 
\hline
&  Non-Hodgkin &  F &  113 &  97.3 &  9.88 &  0.222 &  2.65 \\ 
\hline
&  Larynx &  F &  18.6 &  71.9 &  16.7 &  0.365 &  3.91 \\ 
\hline
&  Liver &  F &  24.9 &  92 &  12.8 &  0.1 &  2.69 \\ 
\hline
&  Melanoma &  F &  19.8 &  102 &  3.19 &  0.226 &  1.7 \\ 
\hline
&  Mesothelioma &  F &  15.6 &  90 &  4.51 &  0.187 &  3.04 \\ 
\hline
&  Myeloma &  F &  49.8 &  92.3 &  7.35 &  0.275 &  2.99 \\ 
\hline
&  Oral &  F &  32.4 &  99.4 &  5.52 &  0.235 &  2.39 \\ 
\hline
&  Ovary &  F &  269 &  96.6 &  9.21 &  0.211 &  2.27 \\ 
\hline
&  Pancreas &  F &  24.3 &  103 &  14 &  0.1 &  2.77 \\ 
\hline
&  Stomach &  F &  140 &  106 &  3.48 &  0.173 &  2.72 \\ 
\hline
&  Thyroid &  F &  20.8 &  80.7 &  23.6 &  0.113 &  1.28 \\ 
\hline
&  Urinary &  F &  45.2 &  98.8 &  5.6 &  0.251 &  2.72 \\ 
\hline

\end{tabular}
\end{ruledtabular}
\end{table}

Expressing the cancer-specific values of $\alpha$ as $r_\alpha \equiv r_\theta/\sqrt\alpha,$ we found that the best fit $r_c$ and $r_\alpha$ values for all cancers depended on cancer type weakly (Table 1). The average $r_c$ value was  $2.6$  and the mean width of the initial distribution was $r_\alpha=0.22.$ The Bayes Information Criterion (BIC) can be used to compare models taking model complexity into account. The BIC for a model with $r_\alpha=0.16$ fixed for all tumor types was ${\rm BIC}_\alpha =1.75 \times 10^{4}$ while that for models with cancer--type specific initial distributions was ${\rm BIC}_0 = 2.092 \times 10^4.$ Some examples are shown in Fig.~1. Thus, notwithstanding the fact that the initial distribution is tissue and cell-type-specific, as proliferation during development is heterogeneous and different tissues undergo tissue-specific replication and apoptosis cycles, the model selected by the BIC is the one with a universal initial distribution. Of course, cell proliferation during growth and development in children and adults are controlled differently, and our results apply only to  tumorigenesis in adults. With two exceptions (a subtype of Hodgkin lymphoma and testicular cancer), the senescence time was very long, on the order of 90 years.  For these two exceptions, the time interval, $w_s,$ over which the somatic stem cells in these tissues stop dividing was much longer.

\begin{figure}
\includegraphics[width=80mm]{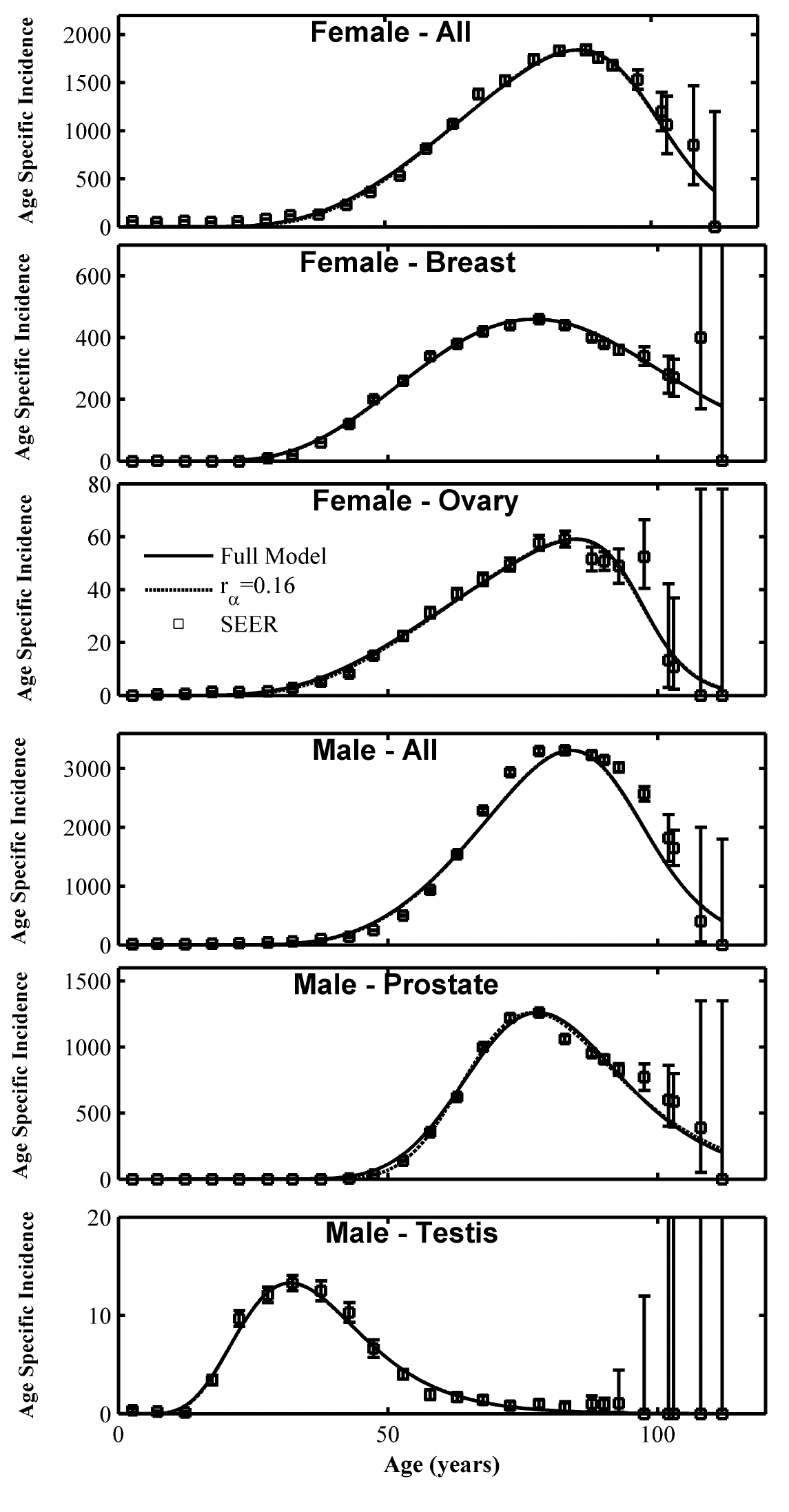}
\caption{ 
Cancer incidence rates. Models with all parameters fitted (Full) and those with $r_\alpha=0.16$ are almost indistinguishable. Incidence is measured per $10^5$ person-years at risk.
}
\end{figure}

We tried to simplify the model further by fixing $r_c$ and $r_\alpha$ for all tumor types and the resulting model with fixed $r_c=4.4$ and $r_\alpha=0.26$ reached a BIC value of ${\rm BIC}_{\alpha,c}=2.087 \times 10^4.$ This value is slightly less than ${\rm BIC}_0$ for the model with $r_c,r_\alpha$ both optimized for each tumor type.  Thus, surprisingly, we find that genomic diffusion, error correction, the initial error distribution, and the critical error radius, are largely universal factors with regards to the incidence of cancer in somatic stem cells in all tissues.  The only factors that are mostly cancer specific are the senescence age and the time interval over which senescence occurs. With a finer categorization of cancer types, and more precise knowledge of the number of somatic stem cells relevant for each cancer type, it should be possible to determine parameters more precisely by including the scale of the coordinate distribution in the optimization, as $r_c$ and $r_\alpha$ are obviously not completely independent.

The inability of some previous models~\cite{ritter40} to match incidence rates that peak and then decrease led Ref.~\cite{hpw} to suggest some possible resolutions. The resolution embodied in our model is that there is a DNA error--correcting process so that every stem cell does not, in fact, proceed to carcinogenesis given enough time, and that senescence correlates the reduction in tumorigenesis with the reduction in mitosis. Indeed, Ref.~\cite{hpw}  explicitly argued that tissue and cellular senescence are the likely biological mechanisms for the observed drop off in cancer incidence in the very elderly. The beta model~\cite{pw}  is also consistent with the latter part of the resolution, but does not have a dynamic basis nor does it posit a role for DNA repair.  

DNA repair involves a multitude of different proteins in distinct pathways~\cite{dnadamage}. It is remarkable that the sum total end result of all these repair mechanisms attempting to correct random mutations is apparently a linear stochastic restoring force. Apart from a constant force, this is the simplest possible functional form. We speculate that a constant restoring force would require an inordinate energetic effort for even innocuous mutations, whereas a linear restoring force allows a graduated escalation in repair effort based on the effective error coordinate. Tolerance to low levels of mutation has also been suggested to be necessary for evolution~\cite{germline}.  This seems to support our linear restoring force, but with the caveat that cancers mostly arise from somatic stem cells whereas evolution is due to germline mutations and the error correction mechanisms need not be the same for the two cell types.  However, it may be the case that a linear restoring force confers higher fitness in both cases although for different reasons.


In conclusion, the simple physical understanding we have presented here suggests that the space of mutational histories has a natural diffusion away from the initial starting distribution, restrained by a universal error correction, and starting from a universal initial distribution. The incidence rate for all cancers is the rate of moving beyond a threshold that depends weakly on tumor type. A relatively sharp demarcation between tumors and normal cells is a concrete prediction of our model of tumorigenesis. This universality suggests that reducing the incidence of sporadic cancers requires enhancing mechanisms that maintain the fidelity of DNA replication in somatic stem cells throughout a lifetime. While this is a facile observation, the universality we found suggests that there are not a multitude of strategies required on a tissue--by--tissue basis. More importantly, there is an interval between $r_\theta=1$ and $r_c\approx 2.6$ for almost all cancers where the mutated somatic stem cell genome has not yet become tumorigenic, and yet may be distinguishable from a normal stem cell, since the initial genome distribution has a width $r_\alpha \approx 0.2$. Detecting somatic stem cells in this interval early, and then targeting therapies towards ablating them, is a possible approach to reducing the incidence of cancer. The universality we have found provides a measure of hope that there may be tissue--independent commonalities in both detection and therapy that could prevent metastases.


\section{Acknowledgement}
This work was supported by the Intramural Research Program of the National Institutes of Health, NIDDK. This study utilized the high-performance computational capabilities of the Biowulf Linux cluster at the National Institutes of Health, Bethesda, MD. (http://biowulf.nih.gov).

{}


\end{document}